\documentclass[12pt,preprint]{aastex}
\usepackage{graphicx}
\usepackage{lscape} 

\slugcomment{Accpeted for publication in The Astronomical Journal;}

\shorttitle{AO Imaging of a Metal-rich Absorber}
\shortauthors{Chun et al.}

\begin{document}


\title{Adaptive Optics Imaging of A Massive Galaxy Associated with a Metal-rich Absorber}


\author{Mark R. Chun} 
\affil{Institute for Astronomy, University of Hawaii, Hilo, HI, 96720}
\author{Varsha P. Kulkarni and Soheila Gharanfoli}
\affil{Dept. of Physics and Astronomy, University of South Carolina, Columbia, SC 29208}
\and
\author{Marianne Takamiya}
\affil{Dept. of Physics and Astronomy, University of Hawaii, Hilo, HI 96720}
\email{}



\begin{abstract}
The damped and sub-damped Lyman-alpha absorption line systems in quasar spectra are believed to be produced by intervening galaxies. However, the connection of quasar absorbers to galaxies is not well-understood, since attempts to image the absorbing galaxies have often failed. While most DLAs 
appear to be metal-poor, a population of metal-rich absorbers, mostly sub-DLAs, has been 
discovered in recent studies. 
Here we report high-resolution K-band imaging with the Keck Laser Guide Star Adaptive Optics 
 system of the field of quasar SDSSJ1323-0021 in search of the galaxy producing the 
$z=0.72$ sub-DLA absorber. With a metallicity of 2-4 times the solar level, this absorber is 
of the most metal-rich systems found to date.

Our data show a large bright galaxy with an angular separation of only 1.25$\arcsec \,$ from the quasar, 
well-resolved from the quasar at the high resolution of our data. 
The galaxy has a magnitude of $K = 17.6-17.9$, which corresponds to 
a luminosity of $\approx 3-6 L^{*}$.  Morphologically, the galaxy is fit with a model 
with an effective radius, enclosing half the total light, of $R_{e} = 4$ kpc and 
a bulge-to-total ratio of 0.4-1.0, indicating a substantial bulge stellar population. 
Based on the mass-metallicity relation of nearby galaxies, the absorber galaxy 
appears to have a stellar mass $\ga 10^{11} \, M_{\odot}$. Given the small impact parameter (9.0  kpc at the absorber redshift), this massive galaxy appears to be responsible for the metal-rich sub-DLA.  The absorber galaxy is consistent with the metallicity-luminosity relation observed for nearby galaxies, but is near the upper end of metallicity. Our 
study marks the first application of laser guide star adaptive optics for study of 
structure of galaxies producing distant quasar absorbers. Finally, this study offers 
the first example of a massive galaxy with a substantial bulge producing a metal-rich absorber.

\end{abstract}





\section{Introduction}

Quasar absorption line systems provide a unique means to probe  the evolution of
interstellar gas to high redshifts ($z \le 6$).  The  damped Lyman-alpha
absorbers (DLAs), with H~I column densities log $N_{\rm H I} > 20.3$ and the
sub-DLAs ($19.0 \le $ log $N_{\rm H I} < 20.3$),  constitute a large fraction of
the neutral gas in galaxies and are thus believed to be  an important  link in
understanding the star formation history of the Universe.   DLAs/subDLAs remain
the only class of galaxies at cosmologically significant redshifts with detailed
measurements of element abundances and can thus trace chemical evolution of
galaxies over $\ge 90 \%$ of the cosmic history. 

In order to fully use the DLA abundance data to study cosmic chemical evolution
in the context of present-day galaxies, it is necessary to combine the abundance
information with information on morphologies, luminosities, colors, and star
formation rates from direct imaging.    However, it has proven hard to obtain
this information for most DLA absorbers and as a result, the type of galaxies
giving rise to the DLAs is far from clear.  DLAs are thus variously thought to
arise in luminous spirals (e.g., Wolfe et al. 1986),   or gas-rich dwarf
galaxies (e.g., York et al. 1986; Matteucci, Molaro, \& Vladilo 1997),  or 
merging proto-galactic fragments in cold dark matter cosmologies (e.g., Haehnelt
et al. 1998), or low surface brightness galaxies (e.g., Jimenez et al. 1999), or
collapsing halos with merging clouds (e.g., McDonald \& Miralda-Escude 1999). 

The lack of substantial chemical evolution found in studies of element
abundances in DLAs (e.g., Kulkarni et al. 2005, P\'eroux et al. 2006b) shows that
the currently known population of DLAs seems to be  dominated by metal-poor
objects. On the other hand, there are  some recent indications that a few
low-$z$ absorbers may be highly enriched--with  near-solar or supersolar
metallicities   (see, e.g., Konig et al. 2006; P\'eroux et al. 2006a; Gharanfoli et al. 2007; 
Meiring et al. 2009b and references therein).  Such metal-rich absorbers could
potentially contribute substantially to the cosmic  metal budget (e.g., P\'eroux
et al. 2006b; York et al. 2006; Prochaska et al. 2006;  Kulkarni et al. 2007;
Khare et al. 2007). It is therefore of special interest to image the  metal-rich
absorbers. 

The primary difficulty in imaging the galaxies producing DLA/sub-DLA absorbers is
the  small angular separation of the galaxy from the (much brighter) background
quasars.  High angular resolution imaging is thus essential to study the stellar
nature of the  absorber galaxies. Adaptive optics (AO) offers a greatly promising
technique to achieve the necessary angular resolution. In an earlier work (Chun et
al. 2006) we reported the first AO imaging study of a few low-redshift DLAs/sub-DLAs,
which uncovered several faint compact objects near the quasars.  Here, we report
results from our Keck laser guide star AO observations of the field of SDSSJ1323-0021, 
a quasar at $z_{em} = 1.4$ with an absorber at $z=0.716$.  

This absorber  was chosen for our study because it is one of the highest metallicity absorbers known.  The 
H I column density was derived by Khare et al. (2004) to be  log $N_{\rm H I} = 20.21^{+0.21}_{-0.18}$ from a Voigt profile fit to the Lyman-$\alpha$ absorption line covered in archival HST UV spectra (from HST program GO 9382, PI Rao).  Khare et al. (2004) reported a metallicity of [Zn/H]=+0.4 based on medium-resolution MMT blue channel spectra. To check this extremely high metallicity value, P\'eroux et al. (2006a) obtained 4.7 km s$^{-1}$ resolution spectra using the Very Large Telescope (VLT)  Ultra-violet Visible Echelle Spectrograph (UVES). Based on multi-component Voigt profile fitting of the metal lines in these high-resolution spectra, P\'eroux et al. (2006a) verified the high metallicity.  Using the $N_{\rm H I}$ value derived by Khare et al. (2004), P\'eroux et al. (2006a) derived [Zn/H] = $+0.61 \pm 0.20$, [Fe/H] = $-0.51 \pm 0.20$, [Cr/H] $< -0.52$, [Mn/H] = $-0.37 \pm 0.20$, and [Ti/H] = $-0.61 \pm 0.22$. 
We note that Prochaska et al. (2006) reported a value of log $N_{\rm H I} = 20.3$ for the same absorber from the same HST data, while Rao et al. (2006) reported log $N_{\rm H I} = 20.54^{+0.16}_{-0.15}$.  
The uncertainty in the $N_{\rm H I}$ value stems from the low S/N in the HST UV spectrum. The 
metallicity estimate of course would be lower if a higher $N_{\rm H I}$ value was adopted. In any case, 
it appears that this absorber is at least 2 times solar in metallicity, and thus one of the most 
metal-rich absorbers known. Furthermore, the large depletions of Fe, Cr, and Ti relative to Zn indicate 
that the absorber is dusty.  Throughout this paper, we adopt the metallicity [Zn/H] = +0.61  for the 
nearly undepleted element Zn derived by P\'eroux et al. (2006a) using log $N_{\rm H I} = 20.21^{+0.21}_{-0.18}$, 
which they point out gives the smallest residual with respect to the HST Lyman-alpha data. 

The AO imaging data presented here were obtained with the goal of detecting and better understanding 
the galaxy underlying the absorber toward SDSSJ1323-0021. The Keck LGSAO system 
offers higher spatial resolution as well as sensitivity in the near-IR compared to previous imaging 
observations of DLAs, and thus offers an unprecedented 
look at a highly interesting absorber galaxy.  Section 2 describes the observations
and data reduction.  Section 3 details the procedure adopted for the subtraction 
of the quasar point spread function. Section 4 describes the results, while section 5  presents a discussion. 
\section{Observations and Data Reduction}

Near-infrared images of the field around SDSSJ1323-0021 
were obtained on
UT 23 May 2007 under photometric conditions with the Keck laser guide star
adaptive optics system (Wizinowich et al. 2006) and the NIRC2 near-infrared camera.  
The narrow camera with a plate scale of $0.0099 \arcsec$ per pixel, and a 
field of view of $10 \arcsec \times 10 \arcsec$ was used. 
106 exposures of 30 second duration each were obtained  in the K\'  \, filter in a standard 9-point 
dither sequence for a total on-source integration of 3180 seconds.  

Standard data reduction techniques for near infrared imaging were used to
produce  the final image. Individual frames were sky-subtracted and
flat-fielded.  Separate  sky frames were constructed for each individual
exposure by averaging source-masked  dithered frames taken within 10 minutes of
the individual exposure.  Flat-fields  were constructed from dome flat images
with the lamps on and off.  Bad pixels were  identified as hot pixels in short
dark frames, as dead pixels in flat frames, and  as pixels with large standard
deviations in either of the dark or flat sequences.

Individual frames were registered to the nearest pixel (10 milliarcseconds)
using the centroid of the quasar and then combined with an average excluding bad
pixels and $\ge 3\sigma$ deviations in the image stack.  The integer pixel
registration is sufficient for these data since in all cases we have at least six
pixels across the FWHM of the images.    In the final mosaiced image, the FWHM 
measured from the quasar is 80 milliarcseconds and the Strehl ratio was estimated to
be between 12\% using the Keck nirc2strehl IDL routine applied to the quasar image itself. 

Object detection was performed using SExtractor (Bertin \& Arnouts 1996) and the quasar subtracted images.  
Images were flux calibrated using observations of 
photometric standard stars taken before and after each quasar.   We estimate a limiting magnitude of 
21.8 magnitude per square arcsecond ($5\sigma$, 1 square arcsec aperture).

\section{Point spread function subtraction}

Detecting and classifying objects close to the line of sight to the quasar in broadband images is limited by our ability to estimate the distribution of light
from the quasar itself.  For images obtained with adaptive optics systems we must contend with a PSF that varies with time.  We have found that 
the most reliable means of obtaining a calibration PSF 
is from in situ measurements from the quasar images themselves.  To explore the line of sight close to the quasar, we performed a two-step 
procedure.  First, we removed the azimuthal average PSF constructed from the final mosaic image.  This azimuthal high-pass filter removes the overall 
profile of the quasar and any component in the image centered on the quasar and azimuthally symmetric.  An important negative consequence
of this is that if the galaxy giving rise to the absorption is centered directly on the quasar line of sight and is
azimuthally symmetric, it will be removed in this process.  In addition to the azimuthal 
average PSF we remove the six-fold and four-fold azimuthally symmetric components of the PSF.  The six-fold pattern arises from diffraction off the 
hexagonal shaped primary mirror and secondary support structure.  The latter causes flares
to sweep out across the image over the course of the observation.  
A four-fold pattern can arise due an alternating piston error in the deformable 
mirror  actuators.  This checkerboard or waffle pattern can lead to a 4-fold pattern from the diffraction off the surface of the mirror.  This "waffle" mode is 
suppressed largely within the system control software (van Dam et al. 2004) and we generally see little evidence of it in the PSFs except in the 
brightest of objects. 

Removing the azimuthal average image leaves a number of positive and negative features in the image within about $0.5"$ of the quasar.  To 
understand if these features are artifacts of the PSF subtraction we project the residual image onto a basis set that describes the temporal
variations of the PSF within the sequence (Chun et al. 2006).  This procedure,
commonly known as a Principle Component Analysis or the Karhunen-Loeve  basis set,
provides a set of eigenvectors describing the PSF variations and the eigenvalues as a
measure of the importance of each eigenfunction in the basis set.  Since we derive the basis set from
the image sequences themselves, the basis set includes noise.  We therefore first perform the azimuthal average
subtraction, then run the object detection algorithm (e.g. SExtractor) to find candidate objects in the field.  Any object
found within $1\arcsec$ of the quasar direction is then subjected to the question of whether it is a result of the PSF subtraction.
The object image is then projected on to the first few terms in our basis set to provide an estimate
of the image content that arises simply due to PSF variations.  If the candidate object has any projection onto any of the first
six eigenmodes, then we remove the object from the detection list.  Inside of a radius of about $0.5"$ the noise 
in the residual image is dominated by the residual from the PSF subtraction.  In the end, there was only one
detected object that satisfied these conditions.

\section{Results}

Fig. 1 shows the deepest $5\arcsec \times 5\arcsec$ region in the final image of the field of SDSSJ1323-0021.  A galaxy (G) 
separated $1.25\arcsec$ from the quasar is clearly detected and corresponds to the object found by SExtractor. 
Fig. 2(a) shows a  $3.3 \arcsec \times 2.2 \arcsec$ section of this image. Fig. 2(b) shows the same 
region after PSF subtraction. Since the residuals within the central 0.5 \arcsec are consistent with being produced in the PSF 
subtraction, we do not consider them real. To obtain a better visual understanding of galaxy G, we show in 
Fig. 2(c) the PSF-subtracted image after 
smoothing by a Gaussian profile with a kernel radius of 5 pixels. Finally Fig. 2(d) 
shows a contour plot of the smoothed PSF-subtracted image from Fig. 2(c).   

The properties of the detected galaxy are listed in Table 1.  While we do not have 
a spectral confirmation of the redshift, given the small angular separation, the chance 
of galaxy G being a random interloper is low. For example, based on Fig. 1(b) of 
Chen \& Lanzetta (2003), the number of random galaxies within a radius of $< 2 \arcsec$ 
with luminosity of $\ga L^{*}$ is $< 0.03$. 
We therefore believe that galaxy G is associated with the absorber. 

At the redshift of the absorber, the angular separation corresponds to a physical impact parameter of 
$9 $ kpc (assuming $H_0 = 70$ km/s/Mpc, $\Omega_M$ = 0.3, and $\Omega_{vac}=0.7$). Both GALFIT (Peng et 
al. 2002) and GIM2D (Simard et al. 2002) were used to analyse the object morphology.  This analysis 
suggests a mostly elliptical profile (GALFIT) with a Sersic n=4.8 to a mixture of bulge and disk 
morphologies with a bulge-to-total ratio of $\sim 0.4$(GIM2D).  The GALFIT profile has an effective radius 
of 62 pixels or 4 kpc if at the redshift of the absorber.  Within this radius the mean signal-to-noise ratio is 1.2 per pixel. 
Our simulations of n=1 and n=4 galaxies input to GALFIT and GIM2D agree with H\"au\ss ler et al. (2007) and find 
that even at  these signal-to-noise ratios the two packages can distinguish gross morphological parameters 
such as the difference between an exponential disk and a pure spheroidal profile.    
 
The apparent magnitude $K' = 17.6$ (GALFIT) corresponds to an absolute magnitude of  
$M_{K} = -25.1$,  -25.0, or -25.0  using k-corrections for E, Sa, or Sc type 
spectral energy distributions, respectively.  The k-corrections were estimated using the IRAF-based cosmopack 
package. (See http://www.iac.es/galeria/balcells/publ\_mbc.html).    Adopting $M^*_K$ from Kochanek et al. (2001) but
converted to the above cosmology ($M^{*}_{K} = -23.5$ for an elliptical and $M^{*}_{K} = -23.0$ for a late-type galaxy) 
the galaxy toward Q1323-0021 would have  a luminosity of  $4.4 L^{*}$ (elliptical) or $6.3 L^{*}$ (late-type spiral). 
For $K' = 17.9$ (GIM2D), the corresponding luminosities are 3.3 $L^{*}$ (elliptical) and 4.8 $L^{*}$ (late-type spiral). 

We note that the morphological parameters are obtained at a mean signal-to-noise ratio
of unity and, in this case, are done within the field of a bright point source.  The measured quasar
brightness is $K'=15.3$ so there is a factor of approximately ten between the integrated brightnesses
of the quasar and the extended galaxy.   The gain brought by the adaptive optics system is
two-fold:  (a) to increase the angular resolution and (b) to dramatically reduce the light in the wings 
of the point spread function.  Together these open the region close to the line of sight for study.

\section{Discussion}

We have detected a large luminous candidate galaxy that is very likely to be the 
sub-DLA absorber toward quasar SDSSJ1323-0021. To put the properties of galaxy G in context, in Fig. 3 we show 
the near-IR luminosity-metallicity relation for galaxies in the Kitt-Peak International Spectrographic Survey (KISS) from 
Salzer et al. (2005). The two panels show two different fits to the near IR luminosity-metallicity 
($L$-$Z$) relation from Salzer et al. (2005), based on different calibrations of the relation between emission-line 
metallicity and the strong-line diagnostic $R_{23}$ . 
The left panel shows the near-IR $L$-$Z$ relation 
derived using the $R_{23}$ calibration of Edmunds \& Pagel (1984)
, while the right panel uses the 
$R_{23}$ calibration of Kennicutt, Bresolin, \& Garnett (2003). 
The larger solid filled circle shows the Zn absorption metallicity of the sub-DLA  toward SDSSJ1323-0021 from P\'eroux et al. (2006a), translated into a corresponding O/H value. We have taken the solar abundance to be log O/H + 12 = 8.66. The absorber toward SDSSJ1323-0021 appears to be at the upper end of the metallicity-luminosity relation. 
If we adopt the alternative K-band $L$-$Z$ relation derived by Salzer et al. (2005) 
with the $R_{23}$ calibration of Kennicutt, Bresolin, \& Garnett (2003), the absorber toward SDSSJ1323-0021 appears to be more extreme compared to the KISS galaxies. 
In either case, the absorber appears to be a 
highly luminous galaxy.  This conclusion is not altered even if the metallicity of the absorber is 
somewhat lower. 

Fig. 4 compares the stellar mass vs. metallicity ($M_{*}$-$Z$) relationship for nearby galaxies from Tremonti et al. (2004) with the metallicity of the absorber toward SDSS1323-0021. Even at the 
upper envelope of the $M_{*}$-$Z$ relation, the metallicity of the absorber (adopted from 
P\'eroux et al. 2006a) implies a stellar mass 
of at least $10^{11} \, M_{\odot}$. (We note, however, that the stellar mass would be lower if the metallicity is lower. For example, if the metallicity is lower by 0.3 dex, the most likely stellar mass would be about $10^{10} \, M_{\odot}$, as inferred from the most probable  $M_{*}$-$Z$ relation.)

We conclude that the supersolar metallicity absorber toward SDSS1323-0021 arises in a massive $L_{*}$ galaxy at an impact parameter of 9 kpc. The morphological detail seen visually in Fig. 2 suggests that the galaxy has a substantial bulge population. 
The large bulge-to-total ratio $0.4 \le B/T \le 1$ indicates a 
classical bulge, rather than a pseudo-bulge (e.g., Gadotti 2009).

One may wonder whether a line of sight passing through a 
galaxy at 9 kpc from the galaxy center can have gas of 
high metallicity. To which we note that the metallicity at the sun's location (8.5 kpc away from the center of the Milky Way) 
is solar. Furthermore, the radial metallicity gradients observed in galaxies are weak, 
and it is not clear whether they even exist at high redshifts, so that the metallicity 
at a galacto-centric distance of 9 kpc need not be much lower than that near the 
center of the galaxy. Finally, if the galaxy G is a  galaxy with a substantial bulge (as seems to be the case from the morphological 
analysis), one might expect it to be substantially metal-rich, and its interstellar material to be fairly uniformly enriched. 

It is interesting to compare how the clues regarding the type of the absorber galaxy 
learnt from our imaging study compare with those derived from element abundance studies. 
Of particular recent interest is the relative abundance of Mn to Fe. 
In the Milky Way, [Mn/Fe] is found to increase with increasing [Fe/H]. This trend could indicate 
that Mn is produced primarily in type Ia supernovae (e.g., Samland 1998). Alternately, the trend 
could arise due to metallicity-dependent yields in type II supernovae 
(e.g., Woosley, \& Weaver 1995; McWilliam et al 2003; Feltzing et al. 2007). Whatever the nucleosynthetic origin of Mn, 
a trend of increasing [Mn/Fe] with increasing metallicity is also seen in DLAs and sub-DLAs (e.g., Meiring et al. 2009a and references therein). 
It is possible for dust depletion effects to mask nucleosynthetic effects (e.g. Kulkarni, Fall, \& Truran 1997; Ledoux, Bergeron, \& Petitjean 2002). However, as discussed in Ledoux et al. (2002) and Meiring et al. (2009a), the [Mn/Fe] trend is likely to reflect nucleosynthetic effects more, because the dust depletion levels of Mn and Fe are roughly similar in warm halo clouds and not too different in warm disk clouds (e.g., Savage \& Sembach 1996; Jenkins 2009).  Ledoux et al. (2002) noted that the [Mn/Fe] vs. [Zn/Fe] trend in DLAs could not be fitted with any dust depletion sequence, even if non-solar values are considered for the intrinsic Mn/Fe ratio, without considering variations of the intrinsic 
[Mn/Fe] ratio.  As noted by Ledoux et al. (2002), the [Mn/Fe] ratio is subsolar for DLAs,  and resembles that in Galactic metal-poor halo stars. For sub-DLAs, the trend of [Mn/Fe] with increasing [Zn/H]  resembles that in Galactic bulge stars (although larger abundance samples are clearly needed to verify this trend). This may suggest that sub-DLAs have an older stellar population that got enriched earlier.   For the sub-DLA toward 
SDSSJ1323-0021, [Mn/Fe] = $+0.14 \pm 0.04$ (P\'eroux et al. 2006a). This [Mn/Fe] value is one of the highest  seen among the absorbers, and lies near the upper right end of the trend in Fig. 11 of Meiring 
et al. (2009a). Such high [Mn/Fe] values 
have been seen in metal-rich Galactic bulge stars (e.g., McWlliam et al. 2003; Alves-Brito et al. 2006). 
Thus, the indication from our AO imaging that the metal-rich absorber toward SDSSJ1323-0021 
arises in a galaxy with a substantial bulge appears to be consistent with the relative abundances in the absorber. 

Our study offers the first example of the LGSAO technique applied to the study of the 
structure of the stellar content of quasar absorber galaxies. The discovery of a 
massive, luminous, galaxy with a substantial bulge in one of the most enriched 
absorbers known is a striking success in finding galaxy counterparts of quasar 
absorbers.  In this case the galaxy is easily detected one arcsecond
from the quasar even at though the galaxy is 2.5 mag fainter and extended.  Our success in detecting a 
galaxy counterpart to the supersolar absorber in 
SDSS1323-0021 is supported by other recent discoveries (though at 
lower resolution) of candidate galaxies producing other metal-rich absorbers (e.g., Straka et al. 2009; P\'eroux et al. 2009). Our results are also consistent with the suggestion that most sub-DLAs (at least the metal-rich ones)  may arise in more 
massive galaxies than DLAs (e.g., Khare et al. 2007; Kulkarni et al. 2009). 

Since the discovery of the metal-rich absorber toward SDSSJ1323-0021, several more absorbers (primarily sub-DLAs) with solar or supersolar metallicities have been found in other recent studies (e.g., Prochaska et al. 2006; Kulkarni et al. 2007, 2009; P\'eroux et al. 2008; Meiring et al. 2009b; and references therein). 
In future, systematic high spatial resolution imaging of more quasar fields with 
absorbers of known metallicities will be critical for understanding how the stellar 
populations of metal-rich and metal-poor galaxies differ. A comparison of the 
images of galaxies underlying DLA and sub-DLA absorbers will also be crucial in 
understanding whether these two populations are distinct, and how they are related. 
Spectroscopic confirmation of the galaxy reported in this study 
should be possible with an integral field spectrograph using AO. 
Unfortunately, the sensitivity of the near-infrared integral field spectrographs currently available is not adequate at the wavelengths of key interesting lines at the redshift of 0.716. 
It will therefore be interesting to obtain confirming 
observations with an AO-supported spectrograph in the future. 
A combination of high-resolution imaging, emission-line spectroscopy, and 
absorption-line spectroscopy for a large number of DLAs/sub-DLAs is essential to fully 
understand where quasar absorbers fit in the big picture of galaxy evolution. 

VPK and SG gratefully acknowledge partial support from the National Science Foundation grant AST-0607739 (PI Kulkarni). VPK is also grateful for partial support from the National Science Foundation grant AST-0908890 (PI Kulkarni). Finally, we thank an anonymous referee for constructive comments that helped to improve the paper. 

\section{References}

Alves-Brito, A., Barbuy, B., \& Allen, D. M. 2006, IAUS, 241, 231

Bertin, E. \& Arnouts, S., 1996, A\&AS, 117, 393.

Chen, H.-W., \& Lanzetta, K. M. 2003, ApJ, 597, 706

Chun, M. R., Gharanfoli, S., Kulkarni, V. P., \& Takamiya, M.  2006, AJ, 131, 686

Feltzing, S., Fohlman, M.,  \& Bensby, T. 2007, A\&A, 467, 665

Gadotti, D. A. 2009, MNRAS, 393, 1531

Gharanfoli, S., Kulkarni, V. P., Chun, M. R., \& Takamiya, M.  2007, AJ, 133, 130

Haehnelt, M. G., Steinmetz, M., \& Rauch, M. 1998, ApJ, 495, 647

H\"aussler, B. et al. 2007, ApJS, 172, 615

Jenkins, E. B. 2009, ApJ, 700, 1299

Junkkarinen, V. T., Cohen, R. D., Beaver, E. A., Burbidge, E. M., Lyons, R. W. \& Madejski, G. 2004, 
ApJ, 614, 658 

Khare, P., Kulkarni, V. P., Lauroesch, J. T., York, D. G., Crotts, A. P. S., \& Nakamura, O. 2004, ApJ, 
616, 86

Khare, P., Kulkarni, V. P., P\'eroux, C., York, D. G., Lauroesch, J. T. \& Meiring, J. D. 2007, 
A\&A, 464, 487 

Konig, B., Schulte-Ladbeck, R. E., \& Cherinka, B.  2006, AJ, 132, 1844

Kulkarni, V. P., Fall, S. M., \& Truran, J. W. 1997, ApJ, 484, L7

Kulkarni, V. P., Khare, P., P\'eroux, C., York, D. G., Lauroesch, J. T. \& Meiring, J. D. 2007, ApJ, 661, 88
Mattucci, Molaro, \& Vladilo 1997 

Kulkarni, V. P., Khare, P., Som, D., Meiring, J. D., York, D. G., P\'eroux, C., \& Lauroesch, J. T. 
2009, MNRAS, submitted

Ledoux, C., Bergeron, J., \& Petitjean, P. 2002, A\&A, 385, 802

McDonald, P.  \& Miralda-Escud\'e,  J.  1999, ApJ, 519, 486

McWilliam, A., Rich, R. M., \& Smecker-Hane, T. A. 2003, ApJ, 592, L21



Meiring, J. D., Kulkarni, V. P., Lauroesch, J. T., P\'eroux, C., Khare, P., \& York, D. G. 2009a, 
MNRAS, 393, 1513

Meiring, J. D., Lauroesch, J. T., Kulkarni, V. P., 
P\'eroux, C., Khare, P., York, D. G., \& Crotts, A. P. S.  2009b, MNRAS, 397, 2037

Peng, Ho, Impey, \&Rix 2002, AJ, 124, 266


P\'eroux, C., Kulkarni, V. P., Meiring, J., Ferlet, R., Khare, P., Lauroesch, J. T., Vladilo, G. \& York, D. G. 2006a, A\&A, 450, 53 
 
P\'eroux, C., Meiring, J. D., Kulkarni, V. P., Ferlet, R., Khare, P., Lauroesch, J. T., Vladilo, G. \& York, D. G.  2006b, MNRAS, 372, 369 

P\'eroux, C., Meiring, J. D., Kulkarni, V. P., 
Khare, P., Lauroesch, J. T., Vladilo, G., \& York, D. G. 2008, MNRAS, 386, 2209

P\'eroux, C., Bouch\'e, N., Kulkarni, V. P., York, D. G., \& Vladilo, G.  2009,  to be submitted
 
Pettini, M. et al. 2000, ApJ, 532, 65

Prochaska J. X., O'Meara J. M., Herbert-Fort S., Burles S., Prochter G., $\&$ Bernstein R., 2006, ApJL, 648, 97

Rao, S. M., Turnshek, D. A., \& Nestor, D. B. 2006, ApJ, 636, 610

Salzer, J., Lee, J. C., Melbourne, J., Hinz, J. L., Alonso-Herrero, A., \& Jangren, A. 2005, ApJ, 624, 661

Samland, M. 1998, ApJ, 496, 155

Savage, B. D., \& Sembach, K. R. 1996, ARAA, 34, 279

Simard, L. et al.  2002, ApJS, 142, 1

Straka, L., Kulkarni, V. P., York, D. G., Woodgate, B. E., \& Grady, C. 2009, AJ, submitted


Tremonti, C.  A. et al. 2004, ApJ, 613, 898 

van Dam, M. A., Le Mignant, D., \& Macintosh, B. A. 2004, SPIE, 5490, 174

Wizinowich, P. L. et al. 2006, PASP, 118, 297

Wolfe, A. M. , Turnshek, D. A., Smith, H. E., \& Cohen, R. D. 1986, ApJS, 61, 249

Woosley, S. E., \& Weaver, T. A. 1995, ApJS, 101, 181

York, D. G.,  Dopita, M., Green, R., \& Bechtold, J.  1986, ApJ, 311, 610

York, D. G. et al.  2006, MNRAS,  367, 945

\clearpage

\begin{figure}
\epsscale{0.75}
\includegraphics{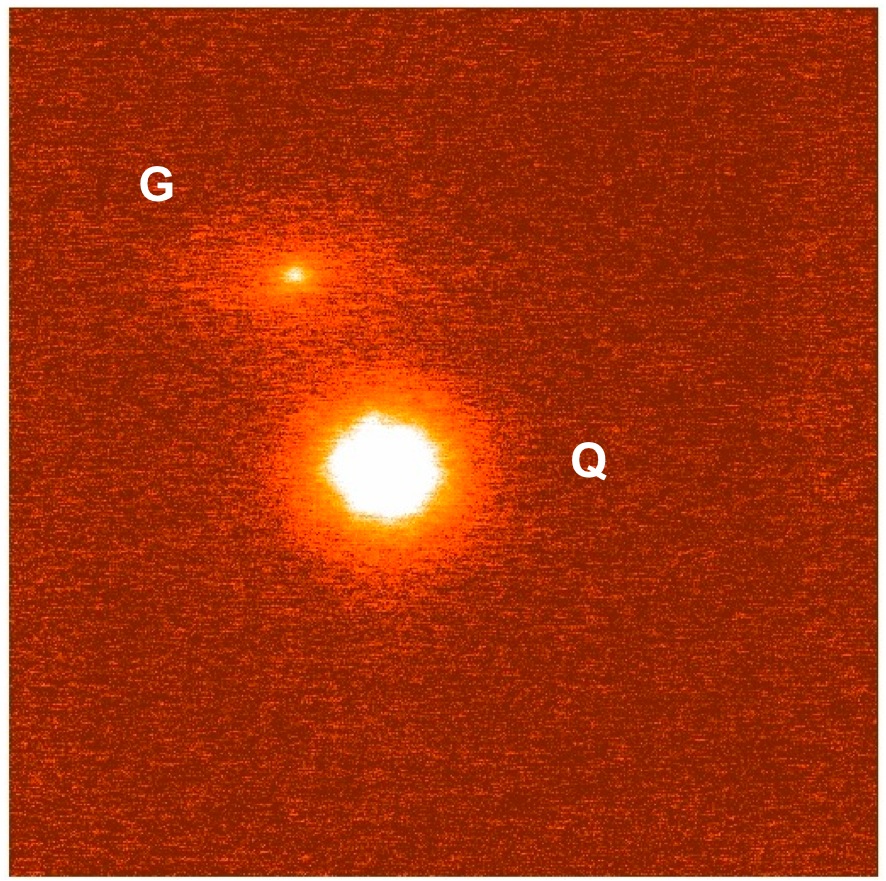}
\caption{Keck Laser Guide Star Adaptive Optics K'-band Image of the field of SDSSJ1323-0021. 
The bright galaxy (``G'') centered at an angular separation of $1.25\arcsec $ from the quasar (``Q'') is the candidate absorber. The field shown 
is $5\arcsec \times 5\arcsec$ and North is up and East is left. }
\label{fig:fig1q1323}
\end{figure} 
      
\begin{figure}
\includegraphics[scale=0.83]{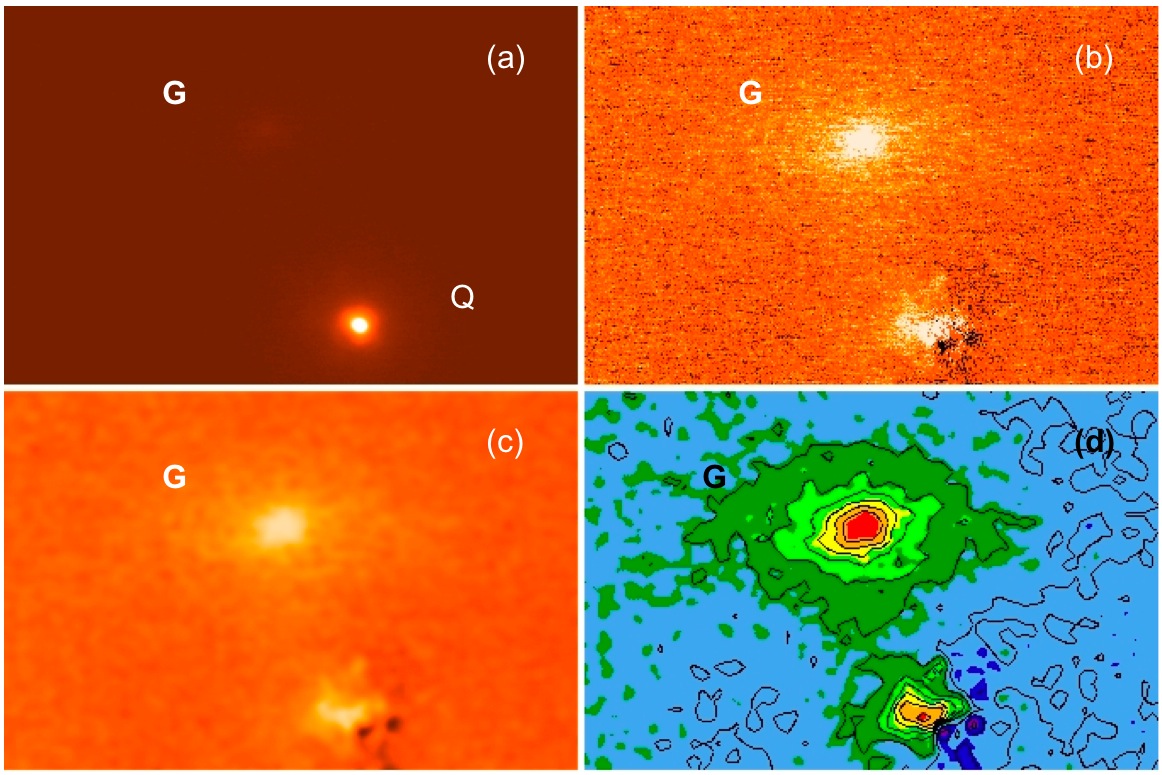}
\caption{$3.3 \arcsec \times 2.2 \arcsec$ regions of the 
Keck Laser Guide Star Adaptive Optics K'-band image of the field of SDSSJ1323-0021.  
(a) unsmoothed image before PSF subtraction; (b) unsmoothed image after PSF subtraction; (c) PSF-subtracted image smoothed by a Gaussian with a kernel radius 
of 5 pixels; (d) a contour map of the PSF-subtracted smoothed image. Based 
on our analysis of intra-night PSF variations, and of images of other point sources, 
we believe that the residuals on top of the quasar left over after PSF subtraction in panels (b), (c), and (d) are artifacts. North is up and East is left in all the panels.}
\label{fig:fig2q1323}
\end{figure} 

\begin{figure}
\includegraphics[scale=0.7]{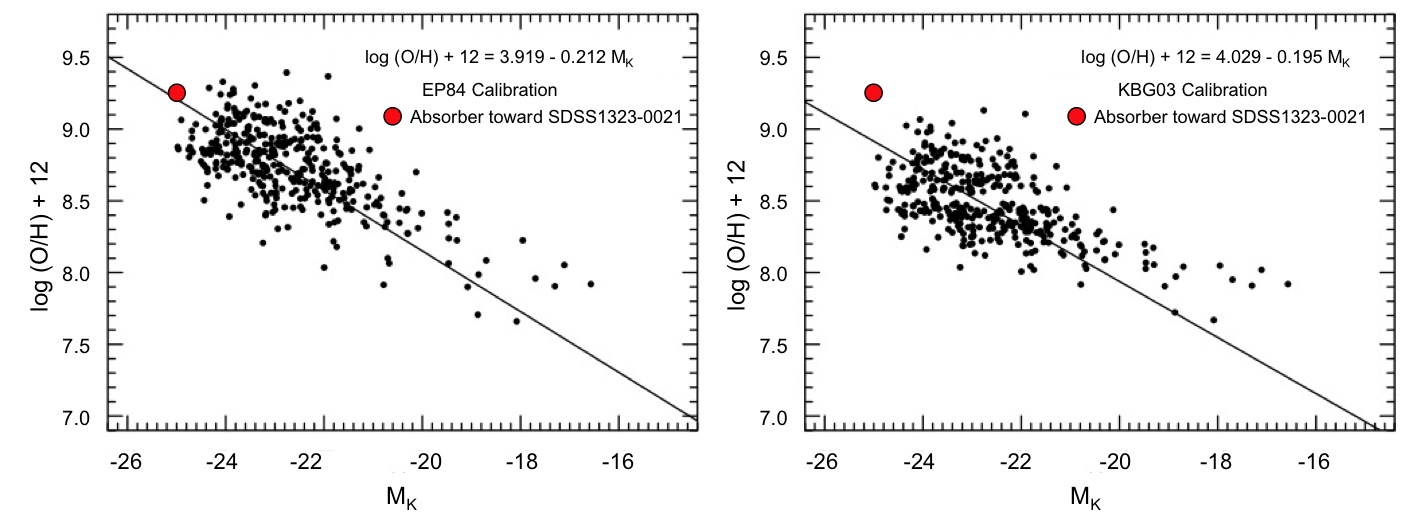}
\caption{A comparison of the candidate absorber galaxy toward SDSS1323-0021 with the near-IR luminosity vs. metallicity relation for galaxies from the KISS survey (Salzer et al. 2005). 
The smaller data points are for 370 star-forming emission-line galaxies from the KISS survey. The line  shows the best-fit to the KISS data for the $R_{23}$ vs. metallicity calibration relation from 
Edmunds \& Pagel (1984).  The larger solid filled circle shows the Zn absorption metallicity of the sub-DLA  toward SDSSJ1323-0021, translated into a corresponding O/H value. }
\label{fig:salzercomp}
\end{figure} 

\begin{figure}
\includegraphics[scale=0.9]{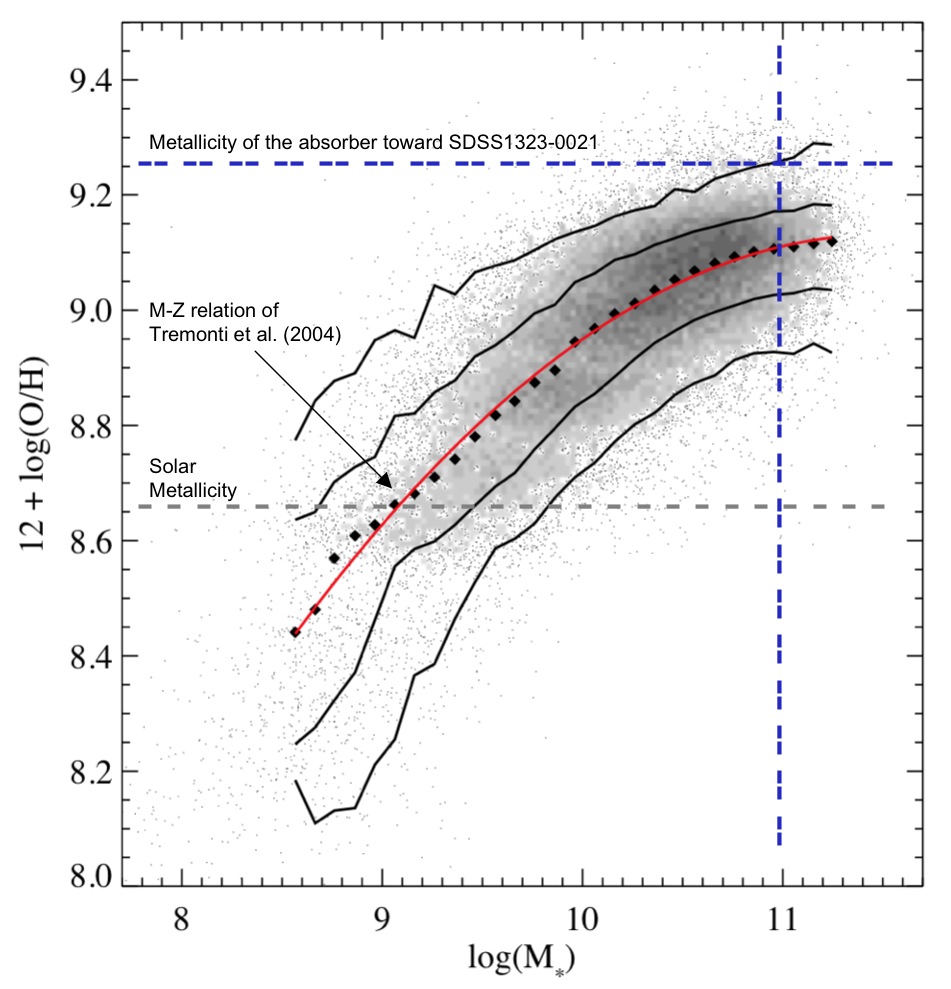}
\caption{A comparison of the metallicity of the absorber galaxy toward SDSS1323-0021 with the 
stellar mass vs. metallicity relation for nearby galaxies from the Sloan Digital Sky Survey (Tremonti et al. 
2004). The black curves are contours enclosing 68\% and 95\% of the data points.  The absorber galaxy toward SDSS1323-0021 appears to have a stellar mass of at least 
$10^{11} M_{\odot}$.}
\label{fig:tremonticomp}
\end{figure} 

\clearpage
\begin{table}
\centerline{\bf{ Table 1}}
\centerline{\bf{Summary of Absorber and Galaxy Properties}}
\vskip15pt
\begin{center}
\begin{tabular}{ll}
\tableline
\tableline
&\\
Absorber Redshift ($z_{abs}$) & 0.716 \\
Absorber H I Column Density (log $N_{\rm H I}$) & 20.21\\
Absorber Metallicity [Zn/H] & +0.61\tablenotemark{a} \\
&\\
Galaxy-Quasar RA separation ($\Delta$ RA) & +0.52 \arcsec \\ 
Galaxy-Quasar Dec. separation ($\Delta$ Dec) & +1.14 \arcsec \\
Total Galaxy-Quasar Angular Separation ($\Delta \theta$) & 1.25 \arcsec\\
Impact Parameter (at $z=0.716$) & 9.0 kpc    \\
Apparent Magnitude ($K$) & 17.6-17.9\tablenotemark{b} \\
Bulge-to-Total Ratio ($B/T$) & $0.4 - 1.0$ \\
Effective Radius ($R_{e}$) & 4.0 kpc        \\
Absolute Magnitude ($M_{K}$) & -24.7 to -25.1 \\
Luminosity & 3.3-6.3 $L^{*}$ \\
Stellar Mass & $\ga 10^{11} M_{\odot}$ \\
&\\
\tableline
\tableline
\tablenotetext{a}{Adopted from P\'eroux et al. 2006a.}
\tablenotetext{b}{Quasar measured apparent brightness is $K = 15.3$.}
\end{tabular}
\end{center}
\label{tab:tabProperties}
\end{table}

\end{document}